# Effect of piezoelectric substrate on phonon-drag thermopower in monolayer graphene

K S Bhargavi[1†], S S Kubakaddi[2*] and C J B Ford[1]

[1]Cavendish Laboratory, Department of Physics, University of Cambridge, Cambridge CB3 0HE, United Kingdom
[2]Department of Physics, K. L. E. Technological University, Hubballi-580031, Karnataka, India
*Email: sskubakaddi@gmail.com

The phonon-drag thermopower is studied in monolayer graphene on a piezoelectric substrate. The phonon-drag contribution $S^g_{PA}$ from the extrinsic potential of piezoelectric surface acoustic (PA) phonons of a piezoelectric substrate (GaAs) is calculated as a function of temperature $T$ and electron concentration $n_s$. At very low temperature, $S^g_{PA}$ is found to be much greater than $S^g_{DA}$ of the intrinsic deformation potential of acoustic (DA) phonons of the graphene. There is a crossover of $S^g_{PA}$ and $S^g_{DA}$ at around ~ 5 K. In graphene samples of about >10 μm size, we predict $S^g$ ~ 20 μV at 10 K, which is much greater than the diffusion component of the thermopower and can be experimentally observed. In the Bloch-Gruneisen (BG) regime $T$ and $n_s$ dependence are, respectively, given by the power laws $S^g_{PA}$ ($S^g_{DA}$) ~ $T^2$($T^3$) and $S^g_{PA}$, $S^g_{DA}$ ~ $n_s^{-1/2}$. The $T$ ($n_s$) dependence is the manifestation of the two-dimensional phonons (Dirac phase of the electrons). The effect of the screening is discussed. Analogous to Herring's law ($S^g\mu_p$ ~ $T^{-1}$), we predict a new relation $S^g\mu_p$ ~ $n_s^0$, where $\mu_p$ is the phonon-limited mobility. We suggest that $n_s$-dependent measurements will play a more significant role in identifying the Dirac phase and the effect of screening.

**PACS:** 72.10.Di, 72.80.Vp, 73.50.Lw, 73.63.-b,

## 1. Introduction

Graphene, a monolayer of carbon atoms bonded in a hexagonal lattice, has been identified as an ideal two-dimensional (2D) system [1-3]. It has attracted intense attention due to its unique linear chiral electronic dispersion with massless 2D Dirac fermions, and it exhibits unique physical properties [4, 5]. Various interesting phenomena such as quantum Hall effect [1,2], ultrahigh mobility [6,7], superior thermal conductivity [8] and high mechanical strength [9] have been observed in graphene. These unusual physical properties make graphene a great potential material for future nanoelectronic devices [4,5].

Since graphene is a one-atom-thick carbon layer, samples are fabricated on substrates for technological applications and fundamental studies. There is an intense search for substrates that improve the quality of electronic properties of monolayer graphene. So far, graphene samples are fabricated on polar substrates such as SiO$_2$, SiC, h-BN, HfO$_2$ and recently GaAs. Most commonly, SiO$_2$ is used as the substrate for graphene devices. But the roughness at the graphene/SiO$_2$ interface and the presence of impurities inducing charge-density fluctuations lead to a reduction in electronic mobility from that observed in suspended graphene. Graphene materials grown on h-BN [10-13] and GaAs [14-17] substrates are expected to give high-quality electronic devices. Transport experiments have shown that graphene on h-BN has higher mobility than has been observed for graphene/SiO$_2$ [10,12]. However, GaAs substrate has advantages over SiO$_2$ and h-BN substrates [14-16]. The substantially larger dielectric constant of GaAs, as compared to SiO$_2$ and h-BN, improves the screening of surface defects and there by increases the mobility of carriers in the graphene layer. Secondly, surface roughness of GaAs is lower than that of SiO$_2$ and this favours a higher quality of the graphene. It is also speculated that the stronger hydrophilic character of GaAs leads to a better stickiness of graphene flakes and thereby prevents their folding. This turned out to make it possible to obtain larger graphene devices with GaAs substrates[16].

In devices with a high-purity GaAs substrate, the electron-phonon (el-ph) interaction can be the decisive factor in limiting the transport properties of Dirac fermions. Therefore, an understanding of the electron-phonon scattering in graphene structures on GaAs substrates is not only of fundamental interest but also of great practical relevance for anticipated new functionalities. The surfaces of such substrates allow for the existence of extrinsic piezoelectric surface acoustic (PA) phonons near the graphene-substrate interface [18,19]. In substrate crystals with lack of a centre of symmetry, such as GaAs, the propagation of a surface acoustic wave induces a piezoelectric potential both inside and outside the GaAs substrate that couples to the electrons in graphene. The Hamiltonian for the interaction of Dirac fermions in graphene with PA phonons due to piezoelectric coupling has been given by Zhang *et al.* [18]. The scattering by PA phonons is shown to play an important role in limiting the electron mobility [18] and electron energy relaxation [19]. The scattering by PA phonons is compared with that due to the potential of intrinsic deformation acoustical (DA) phonons in graphene.

Thermopower is an important transport property, and it is sensitive to both electronic structure and energy dependence of the momentum relaxation time due to various scattering mechanisms. There are two contributions to the thermopower: diffusion $S^d$ and Phonon-drag $S^g$. Phonon-drag thermopower $S^g$ arises due to 'phonon wind', in the temperature gradient, dragging the electrons along with it due to electron-phonon coupling. Consequently, it is found to be purely dependent on electron-acoustic-phonon coupling. Phonon-drag thermopower has been extensively studied in bulk semiconductors [20], the conventional 2DEG in semiconductor heterostructures [21-23] and graphene systems [24], and it is shown to be generally important at low temperatures. Experimental observations in graphene flakes on substrates hundreds of nanometres in lateral size [25-28] show no signature of $S^g$, which is attributed to the weak el-ph coupling. However, theoretical studies in freely suspended monolayer [29, 30] and bilayer [31] graphene of size ~ 10μm considering the electron-acoustic phonon coupling via deformation potential show $S^g$ to be important for temperatures $T \leq 10$ K. In graphene nano-ribbons, $S^g$ is shown to be of order 1 mV/K and sensitive to the Fermi energy and width of the ribbon [32]. We point out that there is a great need for more experimental data on thermopower in graphene below 10 K, especially in the larger samples ( ~ μm). Since most of the graphene devices are on polar substrates, which give rise to piezoelectric surface acoustic phonons, it is important to see the contribution of these phonons to $S^g$. In the present work we investigate the contribution of PA phonons to $S^g$ in monolayer graphene on a GaAs substrate. We expect these calculations to complement the calculations of phonon-limited mobility [18] and hot-electron relaxation [19]. By comparison of the contributions to $S^g$ due to scattering by PA and DA phonons, we show that both mechanisms can be important in different concentration and temperature regimes.

## 2. Basic equations

Here, we consider a gate-controlled monoatomic layer of graphene on a pure GaAs substrate. We take a gate voltage positive enough that the conducting carriers are electrons and their density can be controlled by the applied gate voltage. A carrier in a monolayer of graphene can be described by the two-dimensional (2D) Dirac equation for zero effective mass with the wave function $\Psi_{\lambda \mathbf{k}}(\mathbf{r}) = e^{i\mathbf{k}\cdot\mathbf{r}}[1, \lambda e^{i\theta_\mathbf{k}}]/\sqrt{2}$ and the energy eigenvalue $E_\mathbf{k} = \lambda \hbar v_f |\mathbf{k}|$. Here, $\lambda = +1(-1)$ corresponds to the conduction (valence) band, $v_f = 1\times 10^6$ m/s is the Fermi velocity, $\mathbf{k}$ ($\mathbf{r}$) is the 2D electron wave (position) vector in the x-y plane and $\theta_\mathbf{k}$ is the angle between $\mathbf{k}$ and the x-axis. The corresponding electron velocity $\mathbf{v_k} = (1/\hbar)\nabla_\mathbf{k} E_\mathbf{k}$. The temperature gradient $\nabla T$ is applied in the plane of the layer. This gives rise to an electric field $\mathbf{E} = S/\nabla T$, in the open-circuit condition, where $S = S^d + S^g$ is the thermopower with $S^d$ ($S^g$) being the diffusion (phonon-drag) contribution to the thermopower.

We modify the approach given in Refs. [29, 33] to obtain $S^g$ due to Dirac-fermion coupling with PA phonons. We assume that the 2D Dirac electrons of graphene interact with the 2D surface PA phonons of energy $\hbar\omega_\mathbf{q}$, and the 2D phonon wave vector $\mathbf{q}$. An expression for $S^g$, applicable to both the types of electron-phonon coupling, is shown to be [29]

$$S^g = -\frac{geA}{4\pi^2 \sigma k_B T^2 \hbar^4 v_f} \int_0^\infty dq \int_\gamma^\infty dE_\mathbf{k} \frac{E_\mathbf{k}(\hbar\omega_\mathbf{q})^2 \tau_p \tau(E_\mathbf{k})}{\sqrt{E_\mathbf{k}^2 - \gamma^2}} |C(\mathbf{q})|^2$$
$$\times N_q f(E_k)[1 - f(E_k + \hbar\omega_q)] \quad (1)$$

where $g = g_s g_v$, $g_s = 2$ ($g_v = 2$) is the spin (valley) degeneracy, $A$ is the area of the graphene, $\gamma = \hbar v_f q/2$, $\sigma$ is the electrical conductivity, $\tau_p$ is the phonon relaxation time, $\tau(E_\mathbf{k})$ is the electron relaxation time, $f(E_\mathbf{k})$ is the equilibrium electron distribution, $N_\mathbf{q}$ is the Bose-Einstein distribution for phonons and $|C(\mathbf{q})|^2$ is the square of the electron-phonon matrix element.

The matrix element for the interaction of Dirac electrons of graphene with the piezoelectric surface acoustic phonons of the GaAs substrate is given by $|C(\mathbf{q})|^2_{PA} = [C_{PA}^2 \hbar(e\beta)^2/2A\pi\rho_s v_{sPA}] (q_x q_y/q^2)^2 e^{-2qd} (F(\theta_{\mathbf{k},\mathbf{k'}}))$ [18]. Here, $e\beta = 2.4\times 10^7$ eV/cm is the piezoelectric coupling constant, $\rho_s = 5.3$ g/cm$^3$ is the mass density of the GaAs substrate and $v_{sPA} = 2.7\times 10^5$ cm/s is the velocity of surface acoustic phonons in a GaAs cubic crystal [18], $C_{PA} = 4.9$ is the numerical factor determined by the elastic properties of GaAs and $F(\theta_{\mathbf{k},\mathbf{k'}}) = [1+\cos(\theta_{\mathbf{k},\mathbf{k'}})]/2 = [1 - (q/2k)^2]$ is the overlap integral of the spinor wave functions with $\theta_{\mathbf{k},\mathbf{k'}}$ being the angle between $\mathbf{k}$ and $\mathbf{k'}$. For the distance between the graphene and substrate $d = 5$ Å, $e^{-2qd} \sim 1$ and from the angular average $(q_x q_y/q^2)^2 = \frac{1}{4}$ [18,19]. This makes $|C(\mathbf{q})|^2_{PA}$ nearly independent of $q$, whereas that due to deformation potential coupling varies as $q$. However, we retain $e^{-2qd}$ in the formulation of $S^g$ due to PA phonons.

Assuming $\tau(E_\mathbf{k})$ to vary slowly on the energy surface $E_f$, we take $\tau(E_\mathbf{k}) = \tau(E_f)$. The electrical conductivity of graphene, for the case of elastic scattering, is given by $\sigma = (e^2 E_f \tau(E_f))/\pi \hbar^2$. In the boundary-scattering regime for phonons, $\tau_p = \Lambda/v_{sPA}$, where $\Lambda$ is the phonon mean free path, generally taken to be the smallest (~ few µm) dimension of the sample [34,35]. Substituting the above quantities in Eq.(1), the expression for $S^g_{PA}$ due to PA phonons is shown to be

$$S^g_{PA} = -\frac{g\Lambda C_{PA}^2 (e\beta)^2}{2^5 \pi^2 |e|\rho_s E_f v_f \hbar^3 v_{sPA}^3 k_B T^2} \int_0^\infty d(\hbar\omega_\mathbf{q}) \int_\gamma^\infty dE_\mathbf{k} (\hbar\omega_\mathbf{q})^2 e^{-2qd}$$
$$\times [1-(\gamma/E_k)^2] N_\mathbf{q} f(E_\mathbf{k})[1-f(E_\mathbf{k} + \hbar\omega_\mathbf{q})]. \quad (2)$$

In the Bloch-Gruneisen (BG) regime, $T \ll T_{BGPA}$, defined by a characteristic temperature $k_B T_{BGPA} = 2\hbar v_{sPA} k_f$, $\hbar\omega_\mathbf{q} \sim k_B T$ and $q \ll k_f$. For the surface acoustic phonons of the GaAs substrate $T_{BGPA} = 7.29\sqrt{(n_s/10^{12}\text{cm}^{-2})}$ K. In the ultra-low $T$ region $q \to 0$, and $[1-(\gamma/E_f)^2]^{1/2} \approx 1$ and $[1+(\hbar\omega_\mathbf{q}/E_f)] \approx 1$. In this temperature regime $f(E_\mathbf{k})[1-f(E_\mathbf{k} + \hbar\omega_\mathbf{q})] \approx \hbar\omega_\mathbf{q}(N_\mathbf{q}+1)\delta(E_\mathbf{k}-E_f)$. Then, integration with respect to $E_\mathbf{k}$ gives the following power law

$$S^g_{PABG} = -S^{g0}_{PABG} T^2, \quad (3) \text{ where}$$

$$S^{g0}_{PABG} = \frac{g\Lambda C_{PA}^2 (e\beta)^2 k_B^3 3!\zeta(3)}{2^5 \pi^2 |e|\rho_s E_f v_f \hbar^2 v_{sPA}^3} \quad (4) \text{ Here } \zeta(n) \text{ is the Riemann zeta}$$

function. We find $S^g_{PABG} \sim n_s^{-1/2}$.

For comparison, we give the formula for electron coupling with DA phonons [29],

$$S^g_{DA} = -\frac{gD^2 \Lambda}{2^3 \pi \rho_g |e| E_f k_B T \hbar^3 v_f v_{sDA}^4} \int_0^\infty d(\hbar\omega_q) \int_\gamma^\infty dE_\mathbf{k} (\hbar\omega_q)^3 [1-(\gamma/E_k)^2]^{1/2}$$
$$\times N_q f(e_k)[1 - f(E_k + \hbar\omega_q)] \quad (5)$$

where $D$ is the acoustic deformation potential coupling constant, $\rho_g = 7.6\times 10^{-8}$ g/cm$^2$ is the graphene areal mass density and $v_{sDP} = 2.0\times 10^6$ cm/s is the intrinsic longitudinal DA phonon velocity in graphene. In BG regime, $T \ll T_{DABG} = 54\sqrt{(n_s/10^{12}\text{cm}^{-2})}$, the equation for $S^g$ is given by the simple power law [29]

$$S^g_{DABG} = -S^{g0}_{DABG} T^3 \quad (6)$$

where

$$S^{g0}_{DABG} = -\frac{D^2 \Lambda k_B^4 4!\zeta(4)}{2\pi |e|\rho_g E_f \hbar^3 v_{sDA}^4 v_f} \quad (7)$$

with $S^g_{DPBG} \sim n_s^{-1/2}$.

The diffusion thermopower in monolayer graphene at low temperature is given by the Mott formula, $S^d = -[\pi^2 k_B^2 T (s+1)]/3|e|E_f$ [29]. Here, $s$ is the exponent of energy dependence of the momentum relaxation time $\tau(E_\mathbf{k}) \sim E_\mathbf{k}^s$ and is taken to be 1 for a screened Coulomb potential [36].

## 3. Results and discussion

In order to present the results numerically, we make the choice of parameters $\Lambda$ and $D$ in the following. Since $S^g \sim \Lambda$, it is essential to make reasonable choice of $\Lambda$. Woszczyna et al. [16] have shown that the graphene samples as large as $150 \times 30$ µm$^2$ can be prepared on a GaAs substrate. Thermal conductivity calculations are demonstrated with $\Lambda$ chosen in the range 3-30 µm and the choice of $\Lambda = 5$ µm gives reasonable agreement with the measured thermal conductivity [34,35]. In order to fit the thermal conductivity data, an effective phonon mean free path (in graphene) $\Lambda_{eff} = \Lambda (1+p)/(1-p)$ is used by modulating the smallest dimension of the sample using a specular parameter $p = 0.9$ [34], which enhances $\Lambda$ by a factor of 20. The value of $0 \leq p \leq 1$ is determined by the roughness of the graphene edges. To present our calculations we choose a reasonable value of $\Lambda = 10$ µm.

In the computation of $S^g_{DP}$ it is important to choose a proper value of $D$ as $S^g_{DA} \sim D^2$. A large range of values of $D$ (3-30 eV) has been measured or calculated [37-43]. Hot-electron cooling experiments show $D = 19$ eV is the best fit to the very low-temperature data [38,39] and it is in agreement with the value predicted by us [29]. Also $D = 20$ eV [40], 18 eV [41] and $25\pm 5$ eV [42], 22 eV [43] are being used to explain the electrical transport properties in graphene. In the present calculations we choose a value of $D= 20$ eV.

We restrict our numerical calculations to the low-$T$ regime (0.1-10 K), where boundary scattering alone limits phonon mean free path, and $n_s= 0.5 - 20 \times 10^{12}$ cm$^{-2}$.

### A. Temperature dependence of $S^g$

We show in Fig. 1 $S^g_{PA}$ and $S^g_{DA}$ and their BG regime contributions as a function of temperature for $n_s=1\times10^{12}$ cm$^{-2}$. Both $S^g_{PA}$ and $S^g_{DA}$ increase with $T$ superlinearly. $S^g_{DA}$ increases more rapidly than $S^g_{PA}$. This increase is slower at higher temperature as found in the 2DEG in semiconductor heterostructures [21]. An important result is that $S^g_{PA}$ is much greater than $S^g_{DA}$ over the large range of $T$ considered. For example, at $T=0.1$ (1) K $S^g_{PA}$ is three orders of magnitude (50 times) greater than $S^g_{DA}$. However, in the high temperature region (at about 5 K) $S^g_{PA}$ and $S^g_{DA}$ cross over. In a conventional 2DEG, $S^g_{PA}$ is dominant over $S^g_{DA}$ for $T < 2$ K, owing to the inverse $q$ dependence of the electron-3D acoustic-phonon matrix element via piezoelectric coupling [21,22]. The total $S^g_{DA} + S^g_{PA}$ increases with $T$, with a small depression in the cross-over region of temperature. The combined $S^g_{DA} + S^g_{PA}$ is about 20 µV/K at 10 K, which is about 5 times the diffusion component $S^d$. $S^g$ can be tuned to still higher value by lowering $n_s$ and enhancing $\Lambda$ with good reflecting edges of graphene.

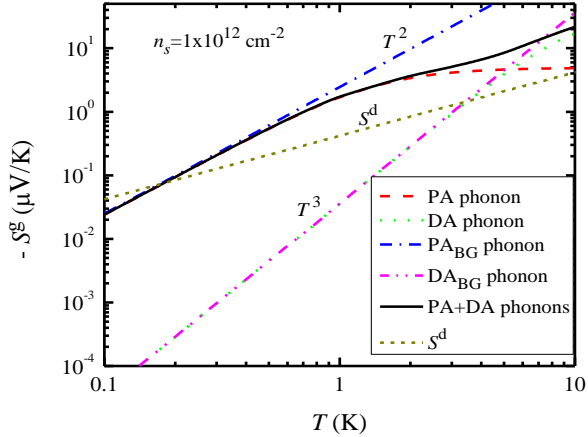

**Figure 1:** Phonon-drag thermopower $S^g$ vs temperature $T$ for both PA and DA phonons for electron concentration $n_s=1\times10^{12}$cm$^{-2}$ along with the corresponding BG regime curves. The diffusion thermopower $S^d$ is also shown for comparison.

The $S^d$ contribution is found to be proportional to $T$ and $n_s^{-1/2}$, and is smaller than the total $S^g$ for the range of temperature $T > 0.2$ K. It gets closer to $S^g$ in the high temperature region However, this relative significance depends upon the choice of $D$ and $\Lambda$.

The BG regime curves are shown by power laws $S^g_{PABG} \sim T^2$ and $S^g_{DABG} \sim T^3$. This difference in exponents is attributed to the difference in the $q$ dependence of the respective el-ph matrix elements. For PA phonon scattering, BG regime is found to be strictly valid for about $T < 0.4$ K and for DA phonon scattering it is valid up to about $T=3$ K. This difference may be attributed to the large difference between $T_{PABG} = 7.311\sqrt{(n_s/10^{12}\text{cm}^{-2})}$ and $T_{DABG} = 54.153\sqrt{(n_s/10^{12}\text{cm}^{-2})}$. $T_{DABG}$ for graphene acoustic phonons is much greater than the $T_{PABG}$ for surface acoustic phonons due to $v_{sPA}$ being about 7.5 times smaller than $v_{sDA}$ in graphene. The temperature dependence of $S^g$ due to unscreened deformation potential coupling $S^g_{DABG} \sim T^4$ in conventional 2DEG [23, 24, 44], and $\sim T^3$ in bilayer graphene [31] and monolayer MoS$_2$ [45] and activated in GNR [32]. These $T^4$ and $T^3$ dependences are, respectively, attributed to the 3D and 2D nature of phonons with linear dispersion, in the respective systems. We also point out that in 3D Dirac-fermion systems, in which phonons are 3D, $S^g_{DABG} \sim T^4$ [46]. For the unscreened piezoelectric coupling $S^g_{PABG} \sim T^2$ for both a conventional 2DEG [23] and a 2DEG in graphene on a GaAs substrate. Although the phonons in conventional 2D systems are taken to be 3D and PA phonons in graphene on a GaAs substrate are 2D, the same $T$ dependence is attributed to the difference in the $q$ dependences of el-ph matrix element. In the former case, the matrix element is $\sim 1/q$ where as in the latter it is nearly independent of $q$ (in the very low-$T$ regime). We may tend to draw the conclusion that the BG regime power-law dependence on $T$ is determined by the dimensionality of the phonons with linear dispersion and the $q$ dependence of the el-ph matrix element. It is independent of the dimensionality of the electron gas and its energy dispersion. Also, it is to be noted that, in BG regime, $S^g_{PABG} \sim (v_{sPA})^{-3}$ and $S^g_{DPBG} \sim (v_{sDA})^{-4}$ and $v_{sDA} \approx 7.5 v_{sPA}$. This is another important reason for the large difference in the magnitudes of $S^g_{PABG}$ and $S^g_{DABG}$ for $T \leq 1$ K.

$S^g_{PA}$ and $S^g_{DA}$ are shown as a function of temperature for different electron concentrations in Figs. 2a and 2b, respectively. Both are found to be smaller for larger $n_s$ at lower $T$ and with the increasing temperature they cross-over so that at higher $T$, $S^g$ is larger for larger $n_s$. This cross-over occurs at around 2 K for $S^g_{PA}$ and 7 K for $S^g_{DA}$. For both PA and DA phonons, the temperature range of validity of the strictly BG regime becomes larger for larger $n_s$. Although, in the low-$T$ region, for a given $T$, the magnitude of $S^g_{PA}$ is larger than that due to $S^g_{DA}$ the crossing over between $S^g_{DA}$ and $S^g_{PA}$ takes place at smaller $T$ for smaller $n_s$ (Fig. 2c).

We notice (from Figs. 2a and 2b) that the temperature dependence can be expressed as $S^g \sim T^{\alpha(T,n_s)}$, where $\alpha(T, n_s)$ is the $T$ and $n_s$ dependent exponent, over the temperature range considered. $\alpha(T,n_s)$ as a function of $T$, for different $n_s$, is shown for PA (DA) phonon coupling in Fig. 3a (3b). For both PA and DA phonon couplings $\alpha(T,n_s)$ is found to decrease from its maximum value 2 to a lower value with increasing $T$ and deviation from 2 begins at higher temperature for larger $n_s$. In this decreasing region of temperature $\alpha(T,n_s)$ is found to be larger for larger $n_s$. We point out that the behaviour of $\alpha(T,n_s)$ is similar to that of the exponent of $T$ in the phonon-limited mobility $\mu_p$ in graphene with the GaAs substrate [18].

### B. Electron concentration dependence of $S^g$

In Fig. 4(a), $S^g$ is shown as a function of $n_s$ (0.5-10$\times10^{12}$ cm$^{-2}$) at $T=1$ K. At this temperature, a general feature is that both $S^g_{PA}$ and $S^g_{DA}$ decrease with increasing $n_s$ and their total is completely dominated by $S^g_{PA}$ over the entire range of $n_s$ considered. For $n_s$ greater than about $4\times10^{12}$ cm$^{-2}$, $S^g_{PA}$ tends to obey BG regime power law $\sim n_s^{-1/2}$ and at lower $n_s$ there is significant deviation from this power law, showing almost an independent behaviour for about $n_s \leq 1\times10^{12}$ cm$^{-2}$. On the other hand, $S^g_{DA}$ obeys the BG-regime power law ($\sim n_s^{-1/2}$) over the entire range of $n_s$ and hence the two curves $S^g_{DA}$ and $S^g_{DABG}$ coincide, as $T_{BG}$ is large in this case even for smaller $n_s$. Also, it is found that $S^d \sim n_s^{-1/2}$. We find that $S^g_{PA}$ is much greater than $S^g_{DA}$ and hence the curve due to total of the two coincides with $S^g_{PA}$. It is to be noted that the BG regime power law $\sim n_s^{-1/2}$, obeyed by both $S^g_{PA}$ and $S^g_{DA}$, is different from the $n_s^{-3/2}$ dependence of these quantities in a conventional 2DEG [23, 44], bilayer graphene [31] and MoS$_2$ [45]. The $n_s^{-3/2}$ dependence in the latter case is attributed to the parabolic dispersion of the electron energy, contrary to the linear dispersion in 2D Dirac fermions.

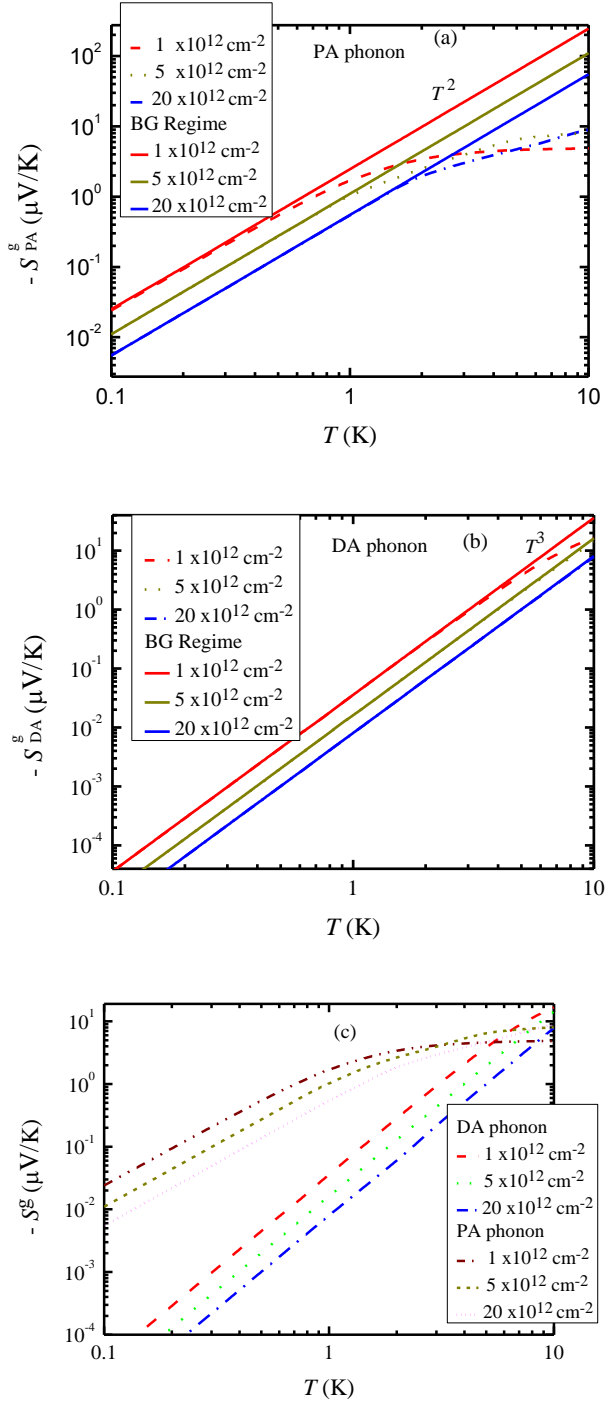

**Figure 2:** Phonon-drag thermopower $S^g$ vs temperature $T$ for the electron concentrations $n_s$=1, 5 and 20 x$10^{12}$ cm$^{-2}$. (a) PA phonons and (b) DA phonons, both with the respective BG-regime curves, (c) Results for PA and DA phonons from (a) and (b), plotted together for comparison.

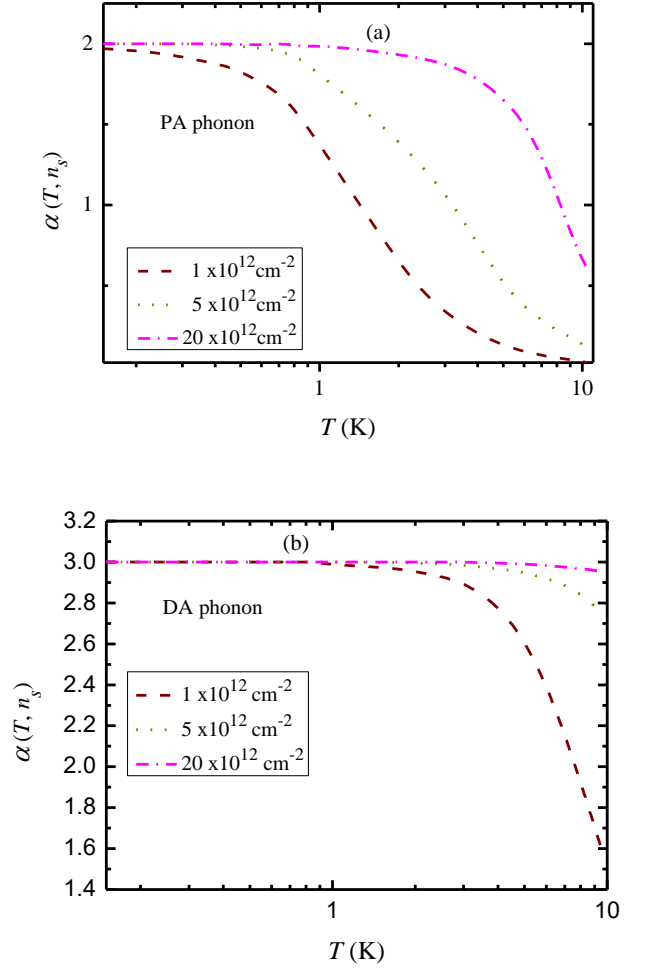

**Figure 3:** Exponent $\alpha(T, n_s)$ vs temperature $T$ for $n_s$= 1, 5 and 20 x$10^{12}$ cm$^{-2}$. (a) PA phonons and (b) DA phonons.

We have shown $S^g$ vs $n_s$ at $T$= 5 K in Fig. 4 (b). Now $S^g_{PA}$ is found to increase with increasing $n_s$, whereas $S^g_{DA}$ is still found to decrease with increasing $n_s$. The total $S^g$ is nearly constant for the $n_s$ range considered. The BG-law curve of $S^g_{PA}$ largely differs from that of the actual $S^g_{PA}$, whereas, the BG-law curve of $S^g_{DA}$ coincides with that of the actual $S^g_{DA}$ for larger $n_s$ and deviates in the low $n_s$ regime.

Expressing $S^g \sim n_s^{-\delta(T,ns)}$, where $\delta(T, n_s)$ is the $T$ and $n_s$ dependent exponent of $n_s$, we have shown $\delta(T,n_s)$ as a function of $n_s$ for different temperatures in Fig. 5. For both PA and DA phonons, $\delta(T,n_s)$ tends towards -0.5 for larger $n_s$. At lower $n_s$, $\delta(T,n_s)$ decreases and changes its sign to positive values for larger $T$. A similar sign change is predicted with the exponent of $n_s$ corresponding to $n_s$ dependence of phonon limited mobility $\mu_p$ [18]. From Figs. 5a and 5b, it is noticed that, for a given temperature, $\delta(T, n_s) = -0.5$ is reached at smaller $n_s$ for DA phonons than for PA phonons.

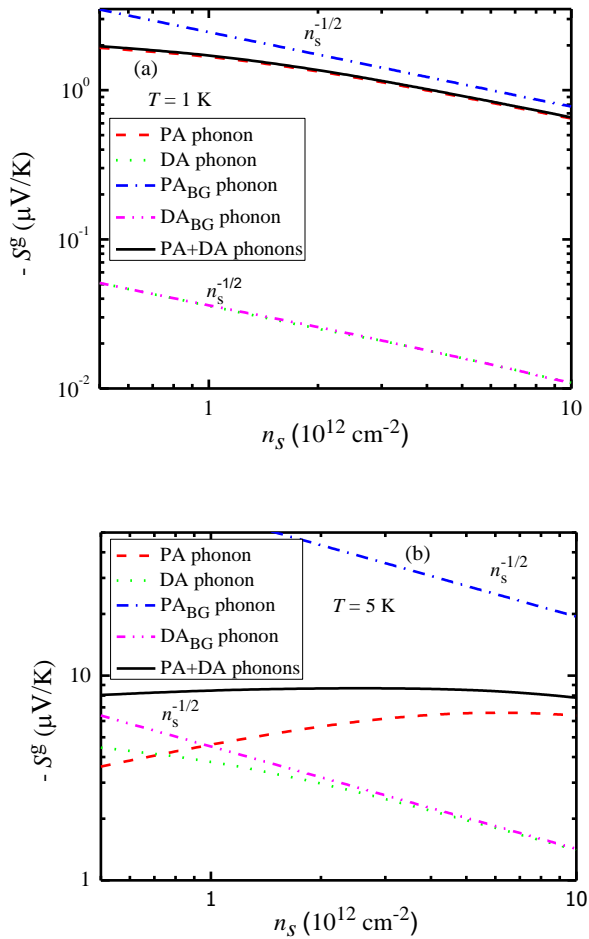

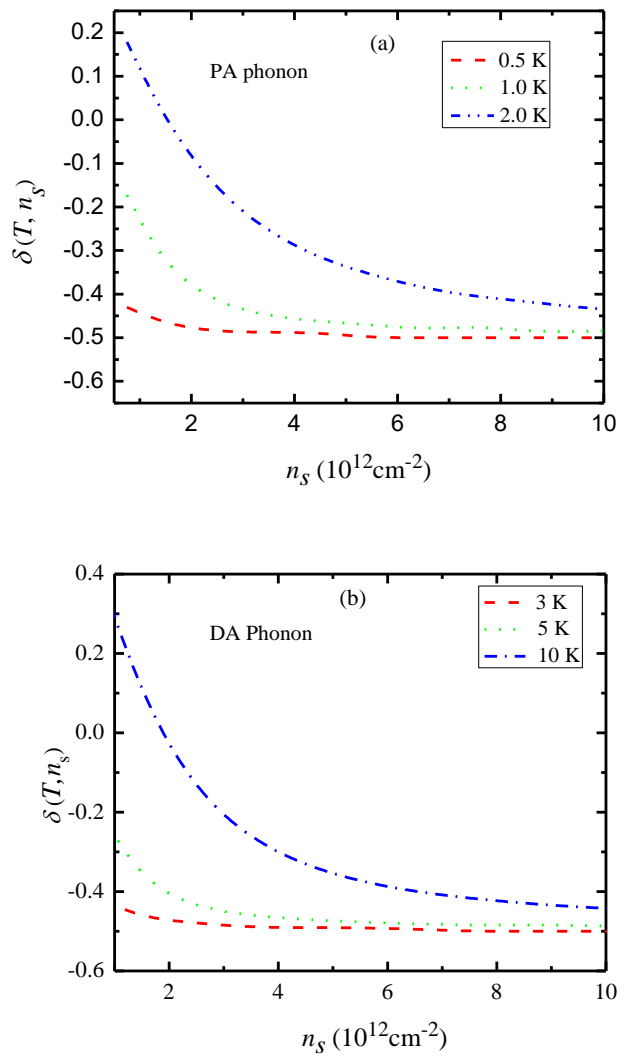

**Figure 4:** Phonon-drag thermopower $S^g$ vs electron concentration $n_s$ along with the respective BG- regime curves. (a) $T$=1 K. Note that $S^g_{DA}$ and $S^g_{DABG}$ curves coincide. (b) $T$=5 K.

### C. Effect of screening

Calculations of the transport properties of graphene exist with and without taking account of screening of el-ph interaction [47]. The need for screening of the el-ph interaction in graphene is not clearly established. In BG regime, the experimental observation of resistivity $\rho \sim T^4$ around 10 K [42], indicates that the screening of el-ph interaction is not strong enough to show $\rho \sim T^6$ dependence. In order to observe the experimentally predicted effect of screening, it is necessary to work in the temperature regime where $\rho$ and the mobility dependence of temperature from non-acoustic phonon mechanisms are unimportant or where unambiguous contributions from these mechanisms can be subtracted. But the properties such as low temperature phonon-drag thermopower and hot-electron cooling, unlike $\rho$, are independent of non-acoustic phonon mechanisms. Experimental observations of hot-electron cooling in BG regime show $T^4$ behaviour [38,39] as predicted from the unscreened electron-acoustic phonon mechanism [29]. However, there is need for more experimental data on thermopower for $T \lesssim 10$ K where $S^g$ is expected to be very significant. In a conventional 2DEG, the effect of screening in the low-$T$ behaviour of $S^g$ has been extensively studied and is well established [21-23].

**Figure 5:** Exponent $\delta(T, n_s)$ vs electron concentration $n_s$ for different $T$. (a) PA phonons and (b) DA phonons.

We present the effect of screening on $S^g$ by dividing the el-ph matrix element by the square of the temperature-independent static dielectric screening function $\varepsilon^2(q) = [1 + (q_{TF}/q)]^2$ (with GaAs substrate), where $q_{TF} = g_s g_v e^2 k_F/\varepsilon_s \hbar v_F$ is the Thomas-Fermi wave vector with $\varepsilon_s = (1+\varepsilon_{GaAs})/2$ [47,48]. In Fig.6a, $S^g$ is shown as a function of $T$ with and without screening for $n_s=1\times10^{12}$ cm$^{-2}$. The screening is found to reduce $S^g$ significantly at low temperature. For eg. at $T=0.1$ K, it is reduced by about $10^3(10^4)$ times for PA(DA) phonon coupling. As $T$ increases, the screening effect decreases. At $T=10$ K the reduction is by about 50 ($10^2$) times for PA (DA) phonon coupling. We notice that effect of screening is more on DA phonon coupling. More importantly, in the BG regime, where $\varepsilon(q) \approx (q_{TF}/q)$, the power law changes from unscreened $S^g_{PABG} \sim T^2$ to screened $S^g_{PABG} \sim T^4$ and $S^g_{DABG} \sim T^3$ to $S^g_{DABG} \sim T^5$. In the inset of Fig. 6a, screened $S^g_{PA}$ and $S^g_{DA}$ are shown as function of $T$ along with the respective BG-regime curves.

The effect of screening with respect to electron concentration is shown in Fig. 6b. $S^g_{PA}$ and $S^g_{DA}$ are shown as a function of $n_s$, at $T=1$ K, with and without screening. Increasing $n_s$ is found to increase screening. This effect is more on $S^g_{DA}$. In the BG regime (inset of Fig. 6b), $n_s$ dependence changes from $n_s^{-1/2}$ to $n_s^{-3/2}$ due to $q_{TF} \sim k_F$. The $n_s^{-1/2}$ dependence of measured hot-electron relaxation in monolayer graphene [39], agreeing with the theoretical prediction [29], may be an indicator of the absence of screening. We suggest that the power laws of the $T$ and $n_s$ dependences of $S^g$ can also be used simultaneously to determine the importance of the screening of the el-ph interaction.

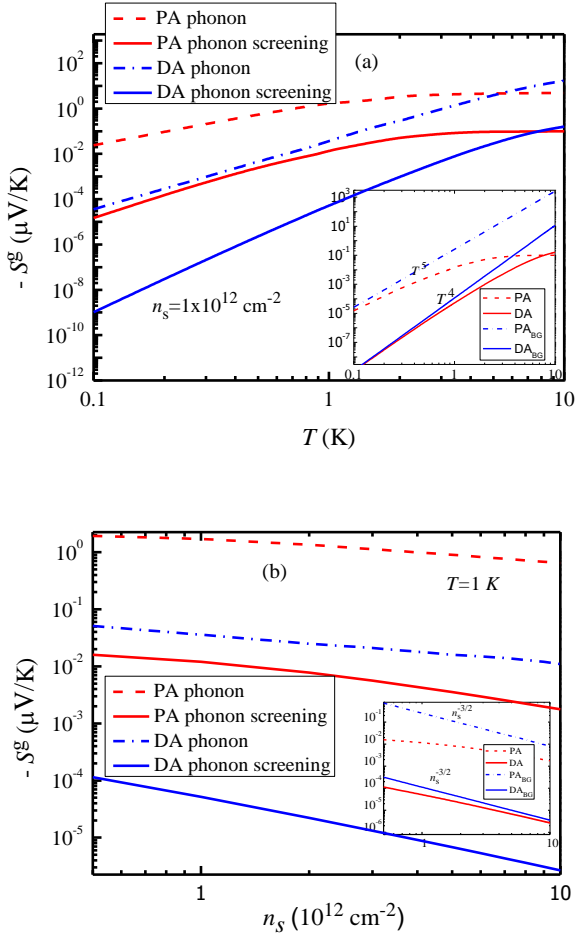

**Figure 6:** Phonon-drag thermopower $S^g$ with and without screening. (a) $S^g$ vs $T$ for $n_s = 1 \times 10^{12}$ cm$^{-2}$ and (b) $S^g$ vs $n_s$ for $T = 1$ K. Insets are screened $S^g$ with the respective BG-regime curves.

### D. Relation of $S^g$ to lattice specific heat $C_v$

In a conventional 2DEG, it has been shown that $S^g \propto f C_v / n_s e$, where $C_v$ is the lattice specific and $f$ is the fraction of momentum lost by the phonons to the carriers [21]. Assuming that $f$ is weakly temperature dependent, then $S^g$ dependence on $T$ comes only from $C_v$. It can be shown that the low temperature 2D lattice specific heat $C_v \sim T^2$ and hence, from this argument, $S^g_{PA} \sim T^2$. Interestingly, at very low temperature, we find $S^g_{PABG} \sim C_v$ contrary to $S^g_{DABG} \sim C_v T$, where $C_v$ in these relations is due to the respective phonons.

### E. Relation of $S^g$ to hot-electron power loss $F(T)$

Since phonon-drag thermopower and low temperature hot electron cooling rate are based on the same basic assumptions and same el-ph coupling with many common factors, we try to find the relation between the two. In the BG regime, we find that the hot-electron cooling rate due to PA phonon coupling is given by $F_{PABG} = F^0_{PABG} T^3$ with $F^0_{PABG} = [g C^2_{PA}(e\beta)^2 k_B^3 2! \zeta(3)]/[2^4 n_s^{1/2} \pi^{5/2} \hbar^3 \rho_s v_f^2 v_{sPA}^2]$ (S S Kubakaddi, unpublished). From this, we find the relation $F_{PABG} = (\xi |e| v_{sPA}/\Lambda) S^g_{PABG} T$. This is similar to the relation obtained in a conventional 2DEG [44] and in graphene for DA phonons [29] with $\xi = 3/2$. This relation can be used to determine one of them if the other is measured.

The dependence of $F_{PABG} \sim n_s^{-1/2}$ is same as that of $S^g_{PABG}$ (Eq.(3)). We notice that, in the BG regime, the hot-electron cooling rate $F_{DABG}$ and $S^g_{DABG}$ due to DA phonon coupling also have the same $n_s^{-1/2}$ dependence [29]. In a conventional 2DEG, with a quadratic dispersion relation, $F_{BG} \sim n_s^{-3/2}$ for both PA and DA phonon coupling [44,49].

### F. Herring's law and a new relation

The acoustic phonon-limited mobility $\mu_p$ and phonon-drag thermopower are related by Herring's law $S^g \mu_p \sim T^{-1}$ in bulk semiconductors [50], conventional 2DEGs [23, 45,51] and graphene [29, 31]. For PA phonons in the present work, using the BG-regime temperature dependence of the phonon-limited mobility $\mu_{pPABG} \sim T^{-3}$ [18], we find that $S^g_{PABG} \mu_{pPA} \sim T^{-1}$ showing that Herring's law is valid for PA phonons. In addition, we find a new relation between these transport coefficients with regard to electron concentration dependence. In the BG regime, $\mu_{pPA} \sim \sqrt{n_s}$ [18] and $S^g_{PABG} \sim n_s^{-1/2}$, so we obtain $S^g_{PABG} \mu_{pPA} \sim n_s^0$ i.e. the product is independent of $n_s$. This is found to be true with DA phonons also in graphene [29]. We have listed in Table I the exponents of the electron concentration dependence, in the BG regime, of $S^g$ and $\mu_p$ in different systems for unscreened DA phonon coupling. Interestingly, we find that this relation $S^g_{DABG} \mu_{pDA} \sim n_s^0$ is true for conventional 2D and 3D semiconductors and their Dirac phases. We expect this relation to be true for PA phonon scattering. Since the screening of the el-ph interaction affects the transport properties in a similar way, we believe this relation to be satisfied for screened $S^g$ and $\mu_p$.

We would like to point out that, the coupling due to in-plane transverse acoustic (TA) phonons is not considered in our present $S^g$ calculations. However, there exist calculations of resistivity $\rho$ [61,62] and hot electron cooling $P$ [63] in which electrons are shown to couple with the TA phonons through an effective gauge field in terms of an unscreened vector potential (VP), besides screened coupling with the DA phonon. In BG regime, $\rho$ [62] and $P$ [63] due to VP coupling are shown to vary as $\sim T^4$, the same as for unscreened DA phonon coupling, where as the screened DA phonon coupling shows $\sim T^6$ dependence [29,62]. Moreover, the contribution due to VP coupling is found to dominate in the low-$T$ regime. Consequently, a generalized in-plane el-ph coupling is introduced by adding the DA and VP couplings [62]. It is shown that replacing $D$ by a fitting parameter $\tilde{D}$ (Eq.48 of Ref. [62]), which is resultant of the screened DA and unscreened VP couplings, will give an excellent agreement with expected results of resistivity [42] for $D \sim 3$ eV and $\tilde{D} \sim 10\text{-}20$ eV. Hence, we believe that, inclusion of VP coupling in our $S^g$ calculations may lead to a choice of new $D$ whose value, in the literature, is in the range 3-30 eV [37-43]. Moreover, we feel that, apart from the $T$ dependence, the more detailed study of the $n_s$ dependence of the above-mentioned transport properties, both experimentally and theoretically, may help to establish the importance of screening. This is because the screening of the DA phonon coupling changes the $n_s$ dependence, in the BG regime, from $\sim n_s^{-1/2}$ to $\sim n_s^{-3/2}$.

### 4. Summary

Phonon-drag thermopower $S^g_{PA}$ due to piezoelectric surface acoustic (PA) phonons of the GaAs substrate is studied in monolayer graphene. At very low temperature, electron- PA phonon coupling is found to be stronger and the $S^g_{PA}$ contribution is very much greater than $S^g_{DA}$ due to intrinsic deformation potential acoustic (DA) phonons of graphene. At higher temperature, $S^g_{DA}$ becomes greater than $S^g_{PA}$. Both $S^g_{PA}$ and $S^g_{DA}$ increase with temperature, but $S^g_{PA}$ increases slowly compared to $S^g_{DA}$. In the Bloch-Gruneisen (BG) regime the power-law is given by $S^g_{PA}$ ($S^g_{DA}$) ~ $T^2$($T^3$), a characteristic of 2D phonons of the respective PA and DA phonon-coupling mechanisms. At very low $T$, both $S^g_{PA}$ and $S^g_{DA}$ depend weakly on electron concentration $n_s$, with the BG-regime power law $S^g_{PA}$, $S^g_{DA}$ ~ $n_s^{-1/2}$, which is a manifestation of 2D Dirac electrons. At higher $T$ both $S^g_{PA}$ and $S^g_{DA}$ increase weakly with increasing $n_s$. This cross-over of weakly decreasing to increasing behaviour takes place at lower (higher) temperature for PA (DA) phonons. Effect of screening of electron-phonon coupling is found to reduce $S^g$ significantly and change the $T$ and $n_s$ dependent power laws in BG regime.

In a result analogous to Herring's law, we have predicted, with regard to the $n_s$ dependence, a new relation $S^g\mu_p$ ~ $n_s^0$, which is found to be valid for conventional and Dirac fermions of 2D and 3D systems. Low-temperature experimental measurements of thermopower in samples of graphene larger than a few μm in size on a GaAs substrate should show the $S^g$ contribution.

**Acknowledgements**

K. S. Bhargavi is grateful for a Dr. D. C. Pavate memorial visiting fellowship at Sidney Sussex College, Cambridge University, U. K.

† Permanent address: Department of Physics, Siddaganga Institute of Technology, Tumkur-572013, Karnataka, India

**References**

1. Novoselov K S, Geim A K, Morozov S V, Jiang D, Zhang Y, Dubonos S V, Grigorieva I V, and Firsov A A, Science **306** 666 (2004).
2. Novoselov K S, Geim A K, Morozov S V, Jiang D, Katsnelson M I, Grigorieva I V, Dubonos S V, and Firsov A A, Nature **438** 197 (2005).
3. Zhang Y B, Tan Y W, Stormer H L, and Kim P, Nature **438** 201 (2005).
4. Neto A H C, Guinea F, Peres N M R, Novoselov K S, and Geim A K, Rev. Mod. Phys. **81** 109 (2009).
5. Das Sarma S, Adam S, Hwang E, and Rossi E, Rev. Mod. Phys. **83** 407 (2011).
6. Morozov S V, Novoselov K, Katsnelson M, Schedin F, Elias D C, Jaszczak J A, and Geim A K, Phys. Rev. Lett. **100** 016602 (2008).
7. Bolotin K I, Sikes K J, Hone J, Stormer H L and Kim P, Phys. Rev. Lett. **101** 096802 (2008).
8. Balandin A, Ghosh S, Bao W, Calizo I, Teweldebrhan D, Miao F, and Lau C N, Nano Lett. **8** 902 (2008); Balandin A, Nat. Mater. **10** 569 (2011).
9. Lee C, Wei X, Kysar J W, Hone J, Science, **321**, 385 (2008).
10. Dean C R, Young A F, Meric I, Lee C, Wang L, Sorgenfrei S, Watanabe K, Taniguchi T, Kim P, Shepard K L and Hone J, Nature Nanotechnol. **5** 722 (2010).
11. Decker R, Wang Y, Brar V W, Regan W, Tsai H-Z, Wu Q, Gannett W, Zettl A, and Crommie M F, Nano Lett. **11** 2291 (2011).
12. Dean C R, Young A F, Cadden-Zimansky P, Wang L, Ren H, Watanabe K, Taniguchi T, Kim P, Hone J and Shepard K L, Nature Phys.**7** 693 (2011).
13. Garcia J M, Wurstbauer U, Levy A, Pfeiffer L N, Pinczuk A, Plaut A. S., Wang L, Dean C R, Buizza R, Van Der Zande A M, Hone J, Watanabe K, and Taniguchi T, Solid State Commun. **152** 975 (2012).
14. Friedemann M, Pierz K, Stosch R, and Ahlers F J, Appl. Phys. Lett. **95** 102103 (2009).
15. Ding F, Ji H, ChenY, Herklotz A, Dörr K, Mei Y, Rastelli A and Schmidt O G, Nano Lett. **10** 3453 (2010).
16. Woszczyna M, Friedemann M, Götz M, Pesel E, Pierz K, Weimann T, Ahlers F J, Appl. Phys. Lett. **100** 164106 (2012).
17. Babichev A V, Gasumyants V E , Yu Egorov A, Vitusevich S and Tchernycheva M Nanotechnol., **25** 335707 (2014).
18. Zhang S H, Xu W, Badalyan S M, and Peeters F M, Phys. Rev. B **87** 075443 (2013).
19. Zhang S H, Xu W, Peeters F M, and Badalyan S M, Phys. Rev. B **89** 195409 (2014).
20. Ure RW, Thermoelectric effects in III–V compounds. In *Semiconductors and Semimetals,* vol. **8** Eds. Willardson. R K and Beer A C, (Academic Press New York,1972), p. 67.
21. Gallagher B L and Butcher P N, in *Handbook on semiconductors* Vol.1 Ed. Landsberg P T (Elsevier, Amsterdam, 1992), p. 817.
22. Fletcher R, Zaremba E and Zeitler U,in *Electron-Phonon interactions in low dimensional structures*, Ed. Challis L., (Oxford Science publications, Oxford 2003), p.149.
23. Tsaousidou M, in *Frontiers in Nanoscience and Nanotechnology*, Vol. **2**, Eds. Narlikar A V and Fu YY (Oxford University Press, Oxford 2010) p. 477.
24. Sankeshwar N S, Kubakaddi S S, and Mulimani B G, in *Graphene Science Handbook: Electrical and Optical Properties*, Edited by Aliofkhazraei M, Ali N, Milne W I, Ozkan C S, Mitura S, Gervasoni J L CRC Hand Book, Vol. **18,** CRC Press, (Taylor and Francis group, New York, 2016), pp. 273.
25. Zuev Y M, Chang W, Kim P, Phys. Rev. Lett. **102** 096807 (2009).
26. Checkelsky J G, and Ong N P, Phys. Rev. B **80** 081413(R) (2009).
27. Wei P, Bao W, Pu Y, Lau C N, Shi J, Phys. Rev. Lett. **102** 166808 (2009).
28. Wu X, Hu Y, Ruan M, Madiomanana N K, Berger C, de Heer W A, Appl. Phys. Lett. **99** 133102 (2011).
29. Kubakaddi S S, Phys. Rev. B **79** 075417 (2009).
30. Bao W S, Liu S Y and Lei X L, J. Phys.: Condens. Matter **22** 315502 (2010).
31. Kubakaddi S S and Bhargavi K S, Phys. Rev. B **82** 155410 ( 2010).
32. Bhargavi K S and Kubakaddi S S, J. Phys.: Condens. Matter **23** 275303 (2011).
33. Cantrell D G and Butcher P N, J. Physics C: Solid State Physics **20,** 1985 (1987); **20** 1993 (1987).
34. Nika D L, Pokatilov E P, Askerov A S, and Balandin A A, Phys. Rev. B **79** 155413 (2009).
35. Ghosh S, Nika D L, Pokatilov E P and Balandin A A, New J. Phys.**11** 095012 (2009).
36. Stauber T, Peres N M R and Guinea F, Phys. Rev. B **76** 205423, (2007); Peres N M R, Lopes dos Santos J M B and Stauber T, Phys. Rev. B **76** 073412 (2007).
37. Borysenko K M, Mullen J T, Barry E A, Paul S, Semenov Y G, Zavada J M,

38. Nardelli M B, and Kim K W, Phys. Rev. B **81** 121412 (R)(2010).
39. Baker A M R, Alexander-Weber J A, Altebaeumer T, Nicholas R J, Phys. Rev. B **85** 5403 (2012).
40. Baker A M R, Alexander-Webber J A, Altebaeumer T, McMullan S D, Janssen T J B M, Tzalenchuk A, Lara-Avila S, Kubatkin S, Yakimova R, Lin C.-T, LiL- J, Nicholas R J, Phys. Rev. B **87** 045414 (2013).
41. Bistritzer R and MacDonald A H, Phys. Rev. B **80** 085109 (2009).
42. DaSilva A M, Zou K, Jain J K, and ZhuJ, Phys. Rev. Lett. **104** 236601 (2010).
43. Efetov D K and Kim P, Phys. Rev. Lett. **105** 256805 (2010).
44. Huang J, Alexander-Webber J A, Janssen T J B M, TzalenchukA, Yager T, Lara-Avila S, Kubatkin S, Myers-Ward R L, Wheeler V D, Gaskill D K, and Nicholas R J, J. Phys.: Condens. Matter **27** 164202 (2015).
45. Fletcher R, Pudalov V M, Feng Y, Tsaousidou M and Butcher P N, Phys. Rev. B **56** 12422 (1997).
46. Bhargavi K S and Kubakaddi S S, J. Phys.: Condens. Matter **26** 485013 (2014).
47. Kubakaddi S S, J. Phys.: Condens. Matter, **27** 455801 (2015).
48. Min H, Hwang E H, and Das Sarma S, Phys. Rev. B **83** 161404(R) (2011).
49. Hwang E H and Das Sarma S, Phys. Rev. B **75** 205418 (2007).
50. Ma Y, Fletcher R, Zaremba E, D'Iorio M, Foxon CT and Harris J J, Phys. Rev. B. **43** 9033 (1991).
51. Herring C, Phys. Rev. **96** 1163 (1954).
52. Tsaousidou M, Butcher P N, and Triberis G P, Phys. Rev. B **64** 165304 (2001).
53. Viljas J K and Heikkila T T, Phys. Rev. B **81** 245404 (2010).
54. Bhargavi K S and Kubakaddi S S, Physica E **56** 123 (2014).
55. Kaasbjerg K, Thygesen K S, and Jauho A-P, Phys. Rev. B **87** 235312 (2013).
56. Kaasbjerg K, Bhargavi K S and Kubakaddi S S, Phys. Rev. B **90** 165436 (2014).
57. Karpus V, Semicond. Sci. Technol., **5** 691 (1990)
58. Das Sarma S, Hwang E H, and Min H, Phys. Rev. B **91** 035201 (2015).
59. Bhargavi K S and Kubakaddi S S, Phys. Status Solidi RRL **10** 248, (2016).
60. Kubakaddi S S (unpublished). We find an explicit expression for $S^g$, in the BG regime, for a degenerate bulk semiconductor $S^g = -(k_B/e)[gm^2D^2(k_BT)^4\tau_p 5!\zeta(5)]/[2^3 3\pi^3\hbar^7 n_v\rho_v v_s^4]$, where $g = g_s = 2$, $n_v$ is the volume concentration of the electrons and $\rho_v$ is the mass density per unit volume.
61. Ridley B K, Quantum Processes in semiconductors, 2$^{nd}$ edition (Oxford science Publications, Clarendon Press 1988) P. 324. In Eq.(8.27) of this book, in the denominator, the electron concentration $n_v$ seems to be missing.
62. Mariani E and von Oppen F, Phys. Rev. B **82**, 195403 (2010).
63. Ochoa H, Castro E V, Katsnelson M I and Guinea F, Phys. Rev. B 83, 235416 (2011).
64. Chen W and Clerk A A, Phys. Rev. B 86 125443 (2012).

**Table I**: In the Bloch-Gruneisen regime, the electron concentration $n$ dependence of $S^g$, $\mu_p$ and $P$ is expressed as $S^g, \mu, P \sim n^\delta$ for unscreened deformation potential coupling. The values of $\delta$ are given in the table below for different systems (with references).

| Property | Monolayer graphene | Bilayer graphene | Monolayer MoS$_2$ | Conventional two-dimensional electron gas | Three-dimensional Dirac semimetal | Bulk semiconductor |
|---|---|---|---|---|---|---|
| $S^g$ | -1/2 [29] | -3/2 [31] | -3/2 [45] | -3/2 [44] | -1/3 [46] | -1 [59] |
| $\mu_p$ | 1/2 [18] | 3/2 [52] | 3/2 [54] | 3/2 [56] | 1/3 [57] | 1 [60] |
| $P$ | -1/2 [29] | -3/2 [53] | -3/2 [55] | -3/2 [44,49] | -1/3 [58] | -1 [60] |